\documentclass[pra,twocolumn,aps,showpacs,nofootinbib]{revtex4-1}

\usepackage{physics}
\usepackage{amsmath}
\usepackage{amssymb}
\usepackage{graphicx}
\usepackage{float}
\graphicspath{{PNG/}}
\usepackage{xcolor}
\usepackage{multirow}
\usepackage[utf8]{inputenc}

\makeatletter
\newcommand*\bigcdot{\mathpalette\bigcdot@{.5}}
\newcommand*\bigcdot@[2]{\mathbin{\vcenter{\hbox{\scalebox{#2}{$\m@th#1\bullet$}}}}}
\makeatother

\begin{document}
\title{Energetics and Efimov states of three interacting bosons and mass-imbalanced fermions in a three-dimensional spherical harmonic trap}

\author{A. D. Kerin}
\author{A. M. Martin}
\affiliation{School of Physics, University of Melbourne, Parkville, VIC 3010, Australia}

\date{\today}

\begin{abstract}
We consider a system of three particles, either three identical bosons or two identical fermions plus an impurity, within a three-dimensional isotropic trap interacting via a contact interaction. Using two approaches, one using an infinite sum of basis states for the wavefunction and the other a closed form wavefunction, we calculate the allowable energy eigenstates of the system as a function of the interaction strength, including the strongly and weakly interacting limits. For the fermionic case this is done while maintaining generality regarding particle masses. We find that the two methods of calculating the spectrum are in excellent agreement in the strongly interacting limit. However the infinite sum approach is unable to uniquely specify the energy of Efimov states, but in the strongly interacting limit there is, to a high degree of accuracy, a correspondence between the three-body parameter required by the boundary condition of the closed form approach and the summation truncation order required by the summation approach. This specification of the energies and wavefunctions forms the basis with which thermodynamic variables such as the virial coefficients or Tan contacts, or dynamic phenomena like quench dynamics can be calculated.
\end{abstract}
\maketitle

\section{Introduction}
\label{sec:Intro}

Ultracold quantum gases provide an ideal testbed for the investigation of many-body quantum systems. The foundation of such studies is the investigation of the properties of few-body systems \cite{busch1998two,fedorov2001regularization, werner2006unitary, kestner2007level, PhysRevA.81.053615, PhysRevA.86.053631, PhysRevA.74.053604,blume2012few}. Through solutions of the few-body problem it is possible to calculate the thermodynamics of many-body quantum gases \cite{PhysRevLett.102.160401, liu2010three, Cui2012,PhysRevA.86.053631,PhysRevLett.96.030401,PhysRevA.85.033634,PhysRevLett.107.030601, mulkerin2012universality,nascimbene2010exploring, Science335_2010, levinsen2017universality} and the quench dynamics of few-body systems \cite{bougas2020stationary,bougas2019analytical,budewig2019quench,kehrberger2018quantum}. Significant research has been undertaken \cite{Cui2012,d2018efimov,jonsell2002universal,blume2010breakdown} into the study of the interacting three-body problem for both Bose and Fermi gases trapped in three-dimensional spherically symmetric harmonic traps. Such work has lead to our understanding of Efimov states \cite{efimov1973energy,efimov1970weakly, jonsell2002universal, kartavtsev2007low, kartavtsev2008universal, endo2011universal, wang2012origin,efimov1971bound} in Bose and mass-imbalanced Fermi systems. 

More specifically, recent theoretical results in quench dynamics allow for the semi-analytic calculation of quench observables in the three-body system for a quench in s-wave scattering length \cite{kerin2022quench, kerin2022effects}. However only the non-interacting to unitary and vice-versa quenches can be considered, unlike the two-body quench where any quench in s-wave scattering length can be considered \cite{kerin2020two}. Generalising the three-body quench quench calculation to arbitrary scattering lengths is then of great interest and the first step towards that is calculating the energy spectrum. This has already been done for equal mass fermions \cite{liu2010three} and here we extend those calculations to the bosonic and mass-imbalanced fermionic cases. In addition to the relevance to quench dynamics this generalisation also allows for the calculation of thermodynamic quantities such as virial coefficients or the Tan contacts \cite{tan2008energetics, tan2008large, tan2008generalized}.

In this work we consider two  three-body systems: (i) a 2+1 fermion system, with mass imbalance between two substrate particles and an impurity particle and (ii) three identical bosons. In each case the particles interact via a contact interaction and are confined to a spherically symmetric harmonic trap. We derive for general interaction strength and, for fermions, arbitrary mass imbalance between the substrate and impurity, the eigenspectrum for the system via a matrix approach \cite{PhysRevLett.102.160401}. For this system we also derive the system eigenstates in the unitary regime, where the s-wave contact interaction dominates. For comparison we revisit the solutions for the three boson system. For the 2+1 Fermi mass imbalanced system, as expected, we find Efimov-like states using the matrix approach. This then leads us to revisit the three-body parameter for investigating Efimov states \cite{jonsell2002universal}. We find a connection between the numeric matrix approach and the three-body parameter approaches in determining that for a given matrix size there is a value of the three-body parameter which produces, to a high degree of accuracy, the same energy spectrum at unitarity. 

\section{Overview of the Three-Body Problem}
\label{sec:Overview}

We consider a system of three interacting particles with arbitrary masses in a three dimensional spherically symmetric harmonic trap. The positions of the three particles are given $\vec{r}_{1}$, $\vec{r}_{2}$, and $\vec{r}_{3}$ with respective masses $m_{1}$, $m_{2}$, and $m_{3}$. The Hamiltonian is given by
\begin{eqnarray}
\hat{H}_{\rm 3b}&=&\sum_{k=1}^{3}-\frac{\hbar^2}{2m_{k}}\nabla_{k}^2+\frac{m_{k}\omega^2 r_{k}^2}{2}, \label{eq:FirstHamiltonian}
\end{eqnarray}
and the interparticle interactions are modelled as contact interactions enforced by the Bethe-Peierls boundary condition \cite{bethe1935quantum}.
For convenience we define the following coordinate transformations, 
\begin{eqnarray}
\vec{C}&=&\frac{m_{1}\vec{r}_{1}+m_{2}\vec{r}_{2}+m_{3}\vec{r}_{3}}{m_{1}+m_{2}+m_{3}}, \\
\vec{r}&=&\vec{r}_{2}-\vec{r}_{1},\\
\vec{\rho}&=&\frac{1}{\gamma}\left( \vec{r}_{3}-\frac{m_{1}\vec{r}_{1}+m_{2}\vec{r}_{2}}{m_{1}+m_{2}} \right),\\
\gamma&=&\sqrt{\frac{m_{1}(m_{1}+m_{2}+m_{3})}{(m_{1}+m_{2})(m_{1}+m_{3})}}.\nonumber 
\end{eqnarray}
We can then rewrite the Hamiltonian as
\begin{eqnarray}
&&\hat{H}_{\rm 3b}=\hat{H}_{\rm COM}+\hat{H}_{\rm rel},\\
&&\hat{H}_{\rm COM}=-\frac{\hbar^2}{2M}\nabla^{2}_{C}+\frac{M\omega^2 C^2}{2},\label{eq:Hcm}\\
&&\hat{H}_{{\rm rel}}=
-\frac{\hbar^2}{2\mu_{12}}\nabla^{2}_{r}+\frac{\mu_{12}\omega^2r^2}{2}
-\frac{\hbar^2}{2\mu_{13}}\nabla^{2}_{\rho}+\frac{\mu_{13}\omega^2 \rho^2}{2}.\quad\label{eq:Hrel}
\end{eqnarray}
where $\mu_{jk}=m_{j}m_{k}/(m_{j}+m_{k})$ and $M=m_{1}+m_{2}+m_{3}$. $\hat{H}_{\rm COM}$ is the centre-of-mass (COM) part of the Hamiltonian, the solution to which is the simple harmonic oscillator (SHO) wavefunction, with mass $M$. $\hat{H}_{\rm rel}$ is the relative part of the Hamiltonian.

Throughout the rest of this paper we consider two specific cases. The first is two identical substrate fermions interacting with a third impurity particle. For this case we set $m_{1}=m_{\rm i}$ and $m_{2}=m_{3}=m$, such that $\mu_{13}=\mu_{12}=\mu=m_{\rm i}m/(m_{\rm i}+m)$. The second case we consider is three identical interacting bosons. In this case $m_{1}=m_{2}=m_{3}=m$ and $\mu_{13}=\mu_{12}=m/2$. For convenience we can in general define $\mu=m_{\rm i}m/(m_{\rm i}+m)$ where $m_{\rm i}=m$ for the bosonic case. Finally we also define the COM harmonic length scale to be $a_{\rm COM}=\sqrt{\hbar/(m_{\rm i}+2m)\omega}$, and the relative harmonic length scale to be $a_{\mu}=\sqrt{\hbar/\mu\omega}$, where for bosons $m_{\rm i}=m$.

For the fermionic case we investigate the effects of mass imbalance and so we define $\kappa=m/m_{\rm i}$. The effects of changing $\kappa$ are investigated in detail in Sections \ref{sec:Summation} and \ref{sec:Hyperspherical}. It should be noted that as the value of $\kappa$ changes so too does $a_{\mu}$. Due to the dependence of $a_{\mu}$ on $\kappa$ we use $a_{m}/a_{\rm s}$ for $\kappa\leq1$ and $a_{m_{\rm i}}/a_{\rm s}$ for $\kappa\geq1$ in Figs. \ref{fig:FermionESpec}, \ref{fig:BosonESpec}, and \ref{fig:ESpeck>1} where $a_{m}=\sqrt{\hbar/m\omega}$ and $a_{m_{\rm i}}=\sqrt{\hbar/m_{\rm i}\omega}$.

In a non-interacting system Eqs. (\ref{eq:Hcm}) and (\ref{eq:Hrel}) can be solved with SHO wavefunctions. However, we impose the Bethe-Peierls boundary condition \cite{bethe1935quantum} to enforce a contact interaction,
\begin{eqnarray}
\lim_{r_{ij}\rightarrow0}\left[\frac{d(r_{ij}\Psi_{\rm 3b})}{dr_{ij}} \frac{1}{r_{ij}\Psi_{3\rm b}}\right]=\frac{-1}{a_{\rm s}}.
\label{eq:BethePeierls}
\end{eqnarray}
Here $\Psi_{\rm 3b}$ is the total three-body wavefunction, $r_{ij}=|\vec{r}_{i}-\vec{r}_{j}|$, and $a_{\rm s}$ is the s-wave scattering length. The COM part of the wavefunction is independent of the boundary condition, and only the relative wavefunction is affected. Enforcing this boundary condition is equivalent to including a Fermi pseudo-potential term in the Hamlitonian  \cite{braaten2006universality}.

If the wavefunction is symmetric under the interchange of particles $j$ and $k$ and satisfies the Bethe-Peierls boundary condition for $r_{ij}$ then it satisfies the Bethe-Peierls boundary condition for $r_{ik}$. This relationship between particle interaction and particle symmetry allows us to consider the 2+1 fermion and three boson cases while specifying only one scattering length. Other cases require additional boundary conditions and are beyond the scope of this work. For example in the 2+1 boson case the scattering length between the two identical bosons is not necessarily the same as between one of those bosons and the third particle, both scattering lengths would need to be specified.

\subsection{General interaction strength}
\label{sec:Summation}
By expressing the relative wavefunction as an expansion over basis states it is possible to obtain the allowable energy states for any value of $a_{\rm s}$ via a matrix eigenvalue problem \cite{PhysRevLett.102.160401, kestner2007level, liu2010three}. Due to the separability of the centre-of-mass and relative Hamiltonians the total three-body wavefunction may be written
\begin{eqnarray}
&\Psi_{\rm 3b}(\vec{C},\vec{r},\vec{\rho})=\chi_{\rm COM}(\vec{C})\psi_{\rm 3b}^{\rm rel}(\vec{r},\vec{\rho}).
\end{eqnarray}
The centre of mass wavefunction, $\chi_{\rm COM}(\vec{C})$, is a SHO wavefunction of lengthscale $a_{\rm COM}$. The relative wavefunction, $\psi_{\rm 3b}^{\rm rel}(\vec{r},\vec{\rho})$, can be written
\begin{align}
\psi_{\rm 3b}^{\rm rel}(\vec{r},\vec{\rho})&=\nonumber\\
(1+\hat{Q})& \sum_{n=0}^{\infty} C_{n}
\psi_{\rm 2b}^{\rm rel}\left(\nu_{nl},\frac{r}{a_{\mu}}\right) R_{nl}\left( \frac{\rho}{a_{\mu}} \right)Y_{lm}(\hat{\rho}),\nonumber\\\label{eq:psi1}
\end{align}
where $\psi_{\rm 2b}^{\rm rel}$ is the relative interacting two-body wavefunctions, first derived by Ref. \cite{busch1998two}, and $\nu_{nl}$ are the energy pseudo-quantum numbers. $R_{nl}$ is the normalisation and radial part of the three dimensional SHO wavefunction and $Y_{lm}$ is the spherical harmonic. $\hat{Q}$ is the symmetrisation operator, it ensures the correct bosonic or fermionic symmetry of the wavefunction. For three identical bosons $\hat{Q}=\hat{P}_{13}+\hat{P}_{23}$, for two identical fermions and one distinct particle $\hat{Q}=-\hat{P}_{23}$, where $\hat{P}_{jk}$ exchanges the location of particles $j$ and $k$. The $C_{n}$ terms are coefficients of expansion. 

The energy of $\psi_{\rm 3b}^{\rm rel}$ is given by
\begin{eqnarray}
E_{\rm rel}=(2\nu_{nl}+	2n+l+3)\hbar\omega,
\end{eqnarray}
where $\nu_{nl}+n$ is constant so that the energy of each term in Eq. (\ref{eq:psi1}) is the same.

The explicit form of the relative two-body wavefunction is \cite{busch1998two}
\begin{eqnarray}
&&\psi_{\rm 2b}^{\rm rel}\left(\nu_{nl},\tilde{r}\right)=N_{\nu_{nl}}e^{-\tilde{r}^2/2}\Gamma(-\nu_{nl})U\left(-\nu_{nl},\frac{3}{2},\tilde{r}^2\right),\\
&&N_{\nu_{nl}}=\sqrt{\frac{\nu_{nl}\Gamma(-\nu_{nl}-1/2)}{2\pi^2a_{\mu}^3\Gamma(1-\nu_{nl})[\psi^{(0)}(-\nu_{nl}-1/2)-\psi^{(0)}(-\nu_{nl})]}},\nonumber
\end{eqnarray}
where $U$ is Kummer's function, $\tilde{r}=r/a_{\mu}$ and $\psi^{(0)}$ is the digamma function of degree 0. The explicit form of $R_{nl}$ is given by
\begin{eqnarray}
R_{nl}\left( \tilde{\rho} \right)&=&N_{nl}\left( \tilde{\rho} \right)^le^{-\tilde{\rho}^2/2}
L_{n}^{l+\frac{1}{2}}\left(\tilde{\rho}^2\right),\\
N_{nl}&=&\sqrt{\sqrt{\frac{1}{4\pi a_{\mu}^6}}\frac{2^{n+l+3}n!}{(2n+2l+1)!!}},\nonumber
\end{eqnarray}
where $L_{n}^{l+\frac{1}{2}}$ is the associated Laguerre polynomial, and $\tilde{\rho}=\rho/a_{\mu}$. The exchange operators are given by
\begin{eqnarray}
\hat{P}_{23}&&f(\vec{C},\vec{r},\vec{\rho})\nonumber\\
&&=f\left(\vec{C},\frac{m\vec{r}}{m+m_{\rm i}}+\gamma\vec{\rho},\frac{(2m+m_{\rm i})m_{\rm i}\vec{r}}{(m+m_{\rm i})^2\gamma}-\frac{m\vec{\rho}}{m+m_{\rm i}}\right),\nonumber\\
\end{eqnarray}
where $m=m_{2}=m_{3}, \quad m_{1}=m_{\rm i}$ and
\begin{eqnarray}
\hat{P}_{13}&&f(\vec{C},\vec{r},\vec{\rho})\nonumber\\
&&=f\left(\vec{C},\frac{m_{i}\vec{r}}{m+m_{\rm i}}-\gamma\vec{\rho},\frac{-(2m+m_{\rm i})m_{\rm i}\vec{r}}{(m+m_{\rm i})^2\gamma}-\frac{m\vec{\rho}}{m+m_{\rm i}}\right),\nonumber\\
\end{eqnarray}
where $m=m_{1}=m_{3}, \quad m_{\rm i}=m_{2}$. In this paper $\hat{P}_{13}$ is only applicable to the bosonic case where $m_{1}=m_{2}=m_{3}=m_{\rm i}=m$.

As per Refs. \cite{PhysRevLett.102.160401, liu2010three} the allowable values of the scattering length and expansion coefficients for a chosen energy (and thus chosen $\nu_{nl}$) can be determined from the following matrix equation,
\begin{align}
\frac{a_{\mu}}{a_{\rm s}}
\begin{bmatrix}
C_{0}\\
C_{1}\\
\vdots
\end{bmatrix}
=
\begin{bmatrix}
X_{00l} & X_{01l} & X_{02l} & \dots\\
X_{10l} & X_{11l} & X_{12l} & \dots\\
\vdots & \vdots & \vdots & \ddots\\
\end{bmatrix}
\begin{bmatrix}
C_{0}\\
C_{1}\\
\vdots
\end{bmatrix},
\label{eq:Matrix}
\end{align}
where
\begin{eqnarray}
X_{n'nl}=
\frac{2\Gamma(-\nu_{n'l})}{\Gamma(-\nu_{n'l}-\frac{1}{2})}\delta_{n'n}
-\eta\frac{(-1)^l}{\sqrt{\pi}}A_{n'nl},
\end{eqnarray}
and
\begin{eqnarray}
A_{n'nl}=\frac{a_{\mu}^3}{N_{\nu_{nl}}}\int_{0}^{\infty} \tilde{\rho}^2R_{n'l}(\tilde{\rho})R_{nl}\left(\frac{\kappa\tilde{\rho}}{1+\kappa}\right)\nonumber\\
\times \psi_{\rm 2b}^{\rm rel}\left(\nu_{nl},\sqrt{\frac{2\kappa+1}{(1+\kappa)^2}}\tilde{\rho}\right) d\tilde{\rho},\label{eq:MatrixIntegral}
\end{eqnarray}
and $\eta=-1$ or $\eta=2$ for the fermionic and bosonic cases respectively. With Eqs. (\ref{eq:Matrix})--(\ref{eq:MatrixIntegral}) the energy spectrum of the relative three-body wavefunction can be calculated for any desired value of $\kappa$. 

Additionally in the $\kappa\rightarrow0$ limit (the heavy-impurity limit) the matrix elements can be determined analytically:
\begin{eqnarray}
A_{n'n,l}=
\begin{cases}
\dfrac{2}{n'-\nu}\sqrt{\dfrac{\Gamma(n+3/2)\Gamma(n'+3/2)}{\pi\Gamma(n+1)\Gamma(n'+1)}} & l=0\\
\quad 0 & l>0
\end{cases}.\quad\quad
\label{eq:AnalyticIntegral}
\end{eqnarray}
However in the heavy substrate limit, $\kappa\rightarrow\infty$, Eq. (\ref{eq:MatrixIntegral}) diverges due to to $\psi_{\rm 2b}^{\rm rel}(\nu,r)$ diverging as $r\rightarrow0$. It should be noted that the $\kappa\rightarrow\infty$ limit can be evaluated using the Born-Oppenheimer approximation \cite{petrov2012few}.

\begin{figure}
\includegraphics[height=5.5cm,width=8.5cm]{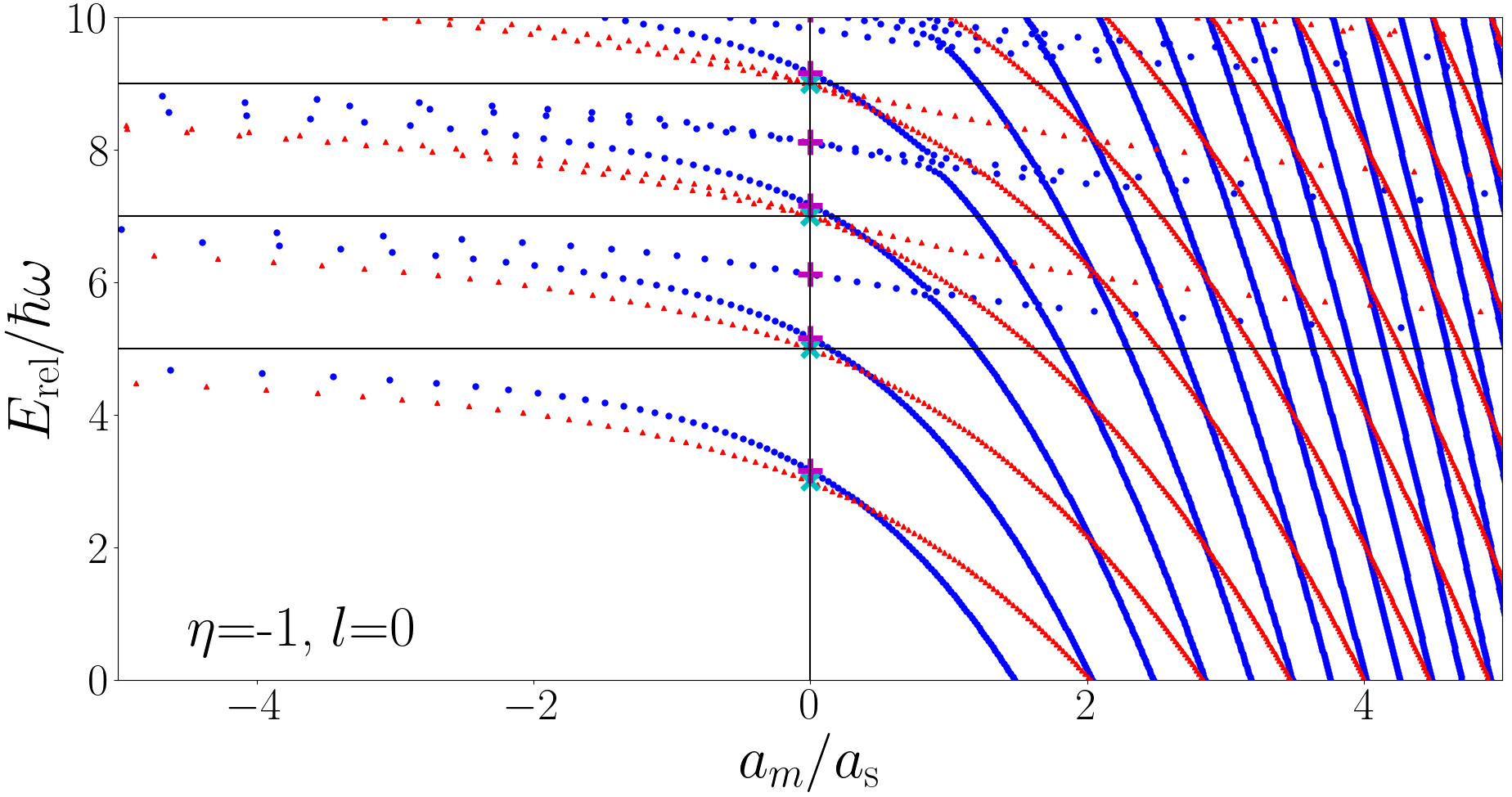}
\caption{Energy spectra of the fermionic three-body relative wavefunction, Eq. (\ref{eq:psi1}), for $l=0$ calculated using Eq. (\ref{eq:Matrix}) with a $50\times50$ matrix. Blue dots correspond to $\kappa=1$ and red upright triangles to the $\kappa\rightarrow0$ limit. The magenta pluses are the unitary energies for $\kappa=1$ and the cyan crosses for $\kappa\rightarrow0$, as per the calculations in Section \ref{sec:Hyperspherical}. The solid horizontal black lines correspond to the non-interacting energies and the solid vertical black line is simply to indicate unitarity.
}
\label{fig:FermionESpec}
\end{figure}

\begin{figure}
\includegraphics[height=5.5cm,width=8.5cm]{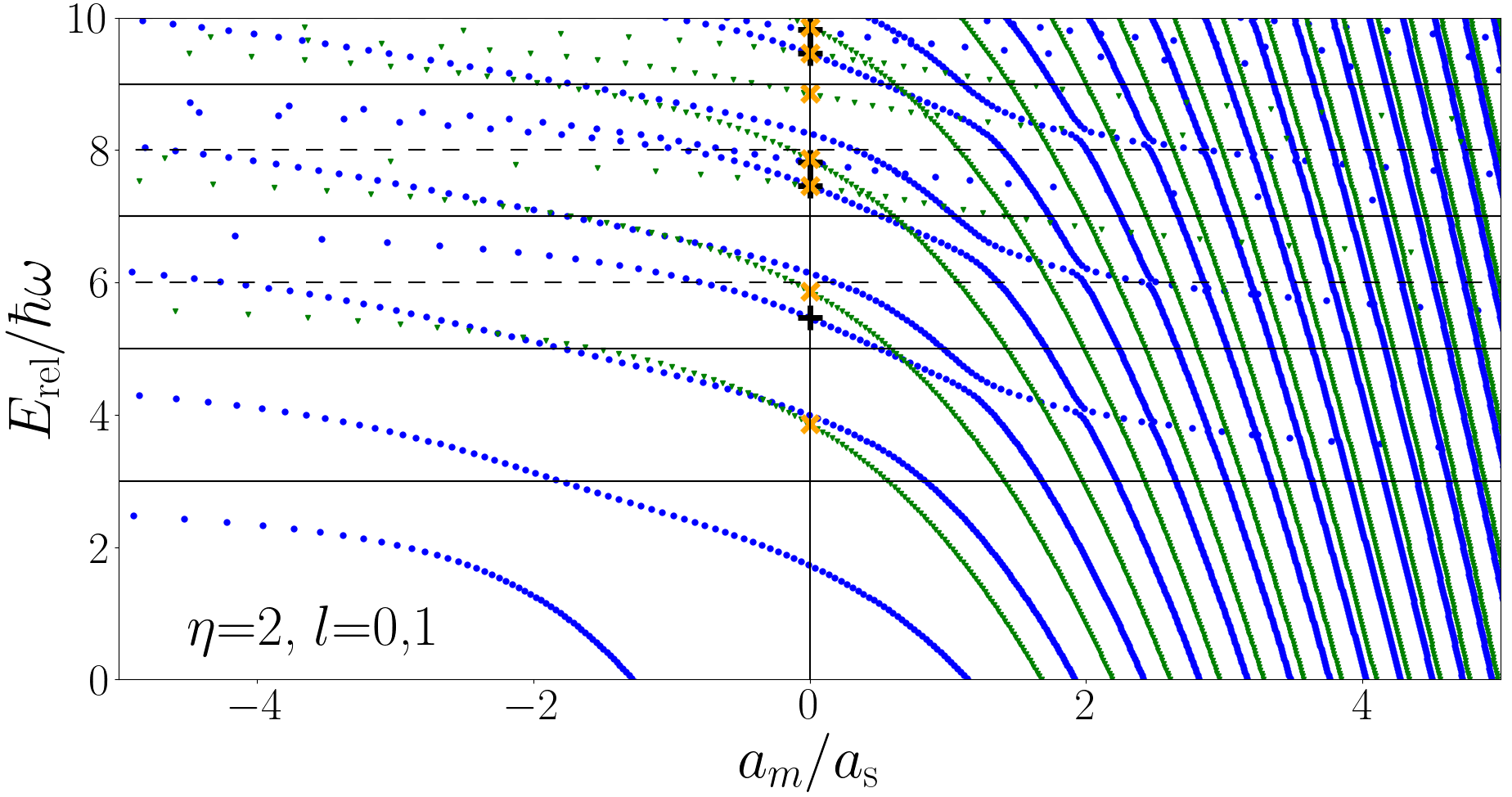}
\caption{Energy spectra of the bosonic three-body relative wavefunction, Eq. (\ref{eq:psi1}), calculated using Eq. (\ref{eq:Matrix}) with a $50\times50$ matrix. Blue dots correspond to $l=0$ and green inverted triangles to $l=1$. The black pluses are the unitary energies for $l=0$ and the orange crosses for $l=1$, as per the calculations in Section \ref{sec:Hyperspherical}. Some $l=0$ unitary energies are unlabelled by a black plus and these correspond to Efimov states as described in Section \ref{sec:Efimov}. The solid horizontal black lines correspond to the non-interacting energies of the $l=0$ state and the dashed horizontal black lines to the $l=1$ state. The solid vertical black line is simply to indicate unitarity.}
\label{fig:BosonESpec}
\end{figure}

\begin{figure}
\includegraphics[height=5.5cm,width=8.5cm]{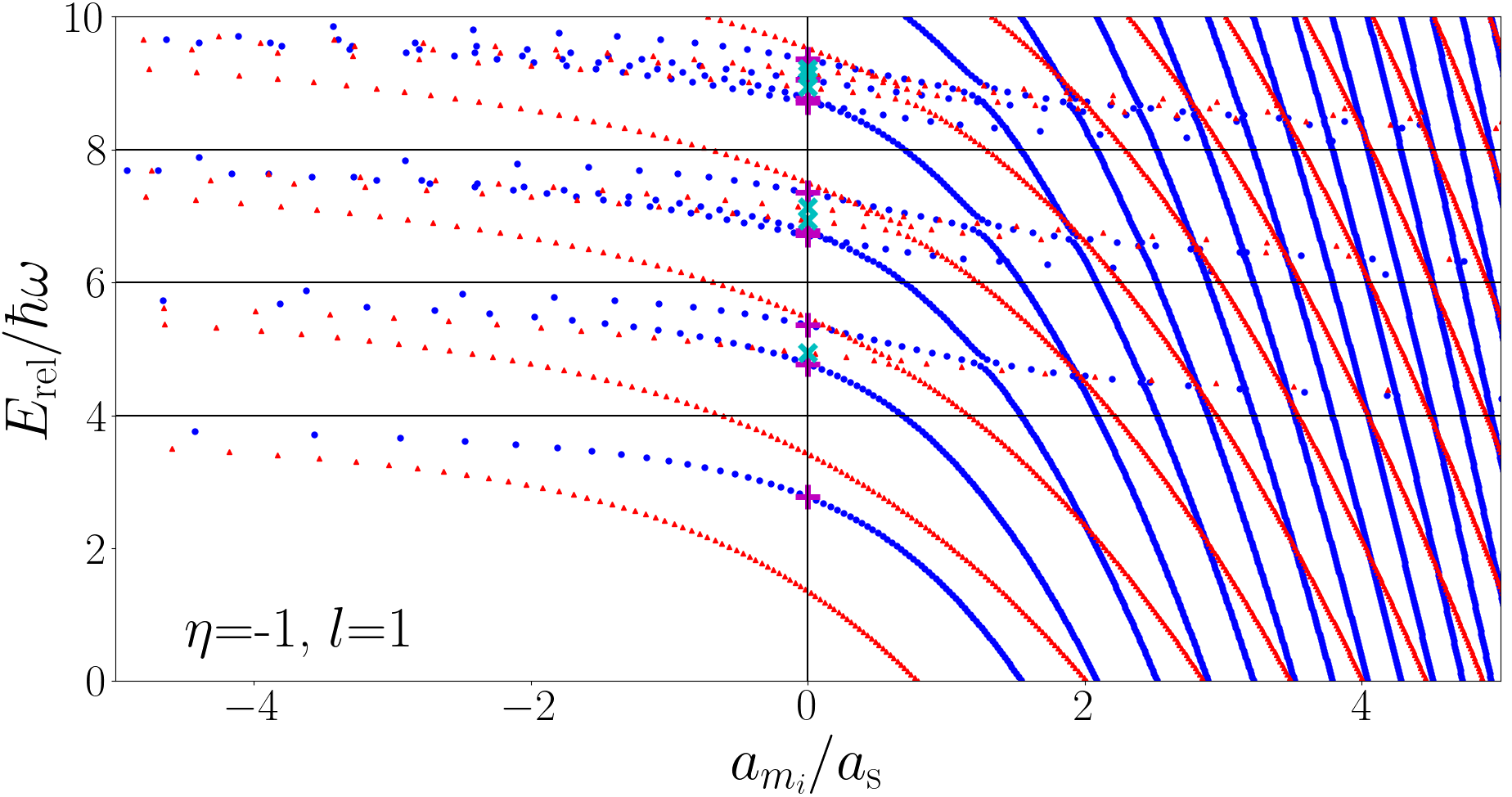}
\caption{Energy spectra of the fermionic three-body relative wavefunction, Eq. (\ref{eq:psi1}), for $l=1$ calculated using Eq. (\ref{eq:Matrix}) with a $50\times50$ matrix. Blue dots correspond to $\kappa=1$ and red upright triangles to $\kappa=13.75$. The magenta pluses are the unitary energies for $\kappa=1$ and the cyan crosses for $\kappa=13.75$, as per the calculations in Section \ref{sec:Hyperspherical}. Some $\kappa=13.75$ unitary energies are unlabelled by a cyan cross and these correspond to Efimov states as described in Section \ref{sec:Efimov}. The solid horizontal black lines correspond to the non-interacting energies and the solid vertical black line is simply to indicate unitarity. }
\label{fig:ESpeck>1}
\end{figure}

The three-body energy spectrum is given in Figs. \ref{fig:FermionESpec}, \ref{fig:BosonESpec}, and \ref{fig:ESpeck>1}. In Fig. \ref{fig:FermionESpec} we present the fermionic energy spectrum for $\kappa \rightarrow 0$ with $l=0$, for comparison, we also show our results for $\kappa=1$, which reproduce the results of Refs. \cite{liu2010three,PhysRevLett.102.160401}. In Fig. \ref{fig:BosonESpec} we present the bosonic energy spectrum for $l=0$ and $l=1$, recall $\kappa=1$ for three identical bosons. In Fig. \ref{fig:ESpeck>1} we present the fermionic energy spectrum for $\kappa=1$ and $\kappa=13.75$ with $l=1$. In all figures the horizontal black lines define the non-interacting energies and the vertical black lines are simply to indicate unitarity $(a_{\rm s}\rightarrow\infty)$. 

In Fig \ref{fig:FermionESpec} we observe for $\kappa=1$ and $a_{\rm s}^{-1}>0$ sharp and close anticrossings are present which make it difficult to clearly identify which unitary states ultimately diverge and which converge as $a_{\rm s}^{-1}\rightarrow+\infty$. However in the $\kappa\rightarrow0$ limit these anticrossings disappear. This makes the identification of which states are ultimately divergent and convergent much more reliable. Notice that for every unitary energy there is one divergent state and the multiplicity at unitarity increases by one every $4\hbar\omega$. This is consistent with Eq. (\ref{eq:kappa0energies}) in Section \ref{sec:Hyperspherical} and implies that the $s_{00}$ states (see Section \ref{sec:Hyperspherical}), and only the $s_{00}$ states, are divergent as $a_{\rm s}\rightarrow+0$. This is similar to findings of Ref. \cite{liu2010three} for the $\kappa=1$ case where the $s_{n=0,l}$ states are identified as the divergent states.

In Fig. \ref{fig:BosonESpec}, the bosonic case, we observe many of the same features as the fermionic spectra in Fig. \ref{fig:FermionESpec}. Again all states converge to non-interacting energies for $a_{\rm s}^{-1} \rightarrow-\infty$ but for $a_{\rm s}^{-1} \rightarrow+\infty$ there are both divergent and convergent states, additionally anticrossings are present. For $l=0$ we are able to identify the $s_{10}$ states as being divergent in the $a_{\rm s}^{-1} \rightarrow+\infty$, however, unlike the $l=0$ fermion spectra discussed above, these are not the only divergent states, all Efimov states (discussed in Section \ref{sec:Efimov}) are divergent in the $a_{\rm s}^{-1} \rightarrow+\infty$ limit. For $l=1$, anticrossings are still present but fewer in number and narrower. The lowest $s_{n=0,l}$ states being divergent no longer appears to be true, the $(q,s_{11})=(0,6.462\dots)$ state is divergent and the $(q,s_{01})=(2,2.864\dots)$ state is not divergent.

In Fig. \ref{fig:ESpeck>1} for $\kappa=13.75$ we again observe anticrossings and that the states corresponding to the smallest real value of $s_{nl}$ diverge as $a_{\rm s}^{-1}\rightarrow +\infty$. However, unlike the bosonic case some, but curiously not all, Efimov states diverge as $a_{\rm s}^{-1}\rightarrow +\infty$. Only the lowest three Efimov energies are divergent in this limit.

\subsection{Hyperspherical approach}
\label{sec:Hyperspherical}
In Section \ref{sec:Summation} we have derived the three-body relative wavefunction for general s-wave scattering length, $a_{\rm s}$, and general mass ratio, $\kappa$, in the form of an infinite sum, Eq. (\ref{eq:psi1}). However, it is possible to derive a closed form eigenfunction of $\hat{H}_{\rm rel}$ using hyperspherical coordinates \cite{werner2006unitary,werner2008trapped}. Only in the unitary and non-interacting cases can the wavefunction be fully specified. Enforcing the Bethe-Peierls boundary condition leads to a transcendental equation that can be solved for one of the quantum numbers, $s_{nl}$, in the non-interacting and unitary regimes. 

We define the hyperradius, $R$, and the hyperangle, $\alpha$,
\begin{eqnarray}
R=\sqrt{r^2+\rho^2}, \quad \alpha=\arctan(\frac{r}{\rho}).
\end{eqnarray}
In turn the relative Hamiltonian is given by
\begin{eqnarray}
\hat{H}_{\rm rel}&=&\frac{-\hbar^2}{2\mu} \Bigg[\frac{\partial^2}{\partial R^2}+\frac{5}{R}\frac{\partial}{\partial R}+\frac{4}{R^2}\nonumber\\
&&+\frac{1}{R^2\sin^2(\alpha)\cos^2(\alpha)}\frac{\partial^2}{\partial \alpha^2}\Big(\cos(\alpha)\sin(\alpha)\bigcdot\Big)\nonumber\\
&&-\frac{\hat{\Lambda}^2_{\rho}}{R^2\cos^2(\alpha)}-\frac{\hat{\Lambda}^2_{r}}{R^2\sin^2(\alpha)} \Bigg]
+\frac{\mu\omega^2 R^2}{2},
\label{eq:HyperHamiltonian}
\end{eqnarray}
where $\hat{\Lambda}^2_{\rho}$ and $\hat{\Lambda}^2_{r}$ are the angular part of the three-dimensional Laplace operators acting on the $\hat{\rho}$ and $\hat{r}$ coordinate spaces respectively. In the unitary and non-interacting limits we can express the wavefunction in closed form. The unnormalised wavefunction is of the form
\begin{eqnarray}
\psi_{\rm 3b}^{\rm rel}&=&\frac{a_{\mu}^2}{R^2}F_{qs_{nl}}(R)(1+\hat{Q})\frac{\varphi_{ls_{nl}}(\alpha)}{\sin(2\alpha)}Y_{lm}(\hat{\rho}).
\label{eq:HyperForm}
\end{eqnarray}
Here $q$, $l$ and $s_{nl}$ are quantum numbers and give the energy
\begin{eqnarray}
E_{\rm rel}=(2q+l+s_{nl}+1)\hbar\omega.
\label{eq:HyperEnergy}
\end{eqnarray}
While $q$ and $l$ are non-negative integers the $s_{nl}$ quantum number is more complicated. 
Following the work of Refs. \cite{werner2006unitary,liu2010three} several conditions can be placed on the wavefunction:
\begin{eqnarray}
\varphi_{ls_{nl}}\left(\frac{\pi}{2}\right)&=&0,\\
s^2\varphi_{ls_{nl}}(\alpha)&=&-\varphi_{ls_{nl}}''(\alpha)+\frac{l(l+1)}{\cos^2(\alpha)}\varphi_{ls_{nl}}(\alpha),\\
E_{\rm rel}F_{qs_{nl}}(R)&=&\frac{-\hbar^2}{2\mu}\left(F_{qs_{nl}}''(R)+\frac{F_{qs_{nl}}'(R)}{R}\right)\nonumber\\
&\quad &+\left(\frac{\hbar^2s_{nl}^2}{2\mu R^2}+\frac{\mu\omega^2 R^2}{2}\right)F(R).\label{eq:HyperradialCondition}
\end{eqnarray}
The first is enforced because a divergence at $\alpha=\pi/2$ is non-physical, the second and third come from requiring that Eq. (\ref{eq:HyperForm}) is an eigenfunction of Eq. (\ref{eq:HyperHamiltonian}). These conditions determine the functional form of the wavefunction. The hyperspherical solutions are given \cite{liu2010three}
\begin{eqnarray}
&&\varphi_{ls_{nl}}(\alpha)=\nonumber\\
&&\cos^{l+1}(\alpha){}_{2}F_{1}\left(\frac{l+1-s_{nl}}{2},\frac{l+1+s_{nl}}{2};l+\frac{3}{2};\cos^2(\alpha)\right),\nonumber\\\label{eq:HyperangularWavefunc}\\
F&&(R)=\left(\frac{R}{a_{\mu}}\right)^{s_{nl}} e^{-R^2/2a_{\mu}^2}L_{q}^{s_{nl}}\left(\frac{R^2}{a_{\mu}^2}\right),\label{eq:HyperradialWavefunc}
\end{eqnarray}
where ${}_{2}F_{1}$ is the Gaussian hypergeometric function. This solution for $F(R)$ is valid for $s_{nl}^2\geq0$. For $s_{nl}^2<0$ there is a different solution, see Section \ref{sec:Efimov}. Note that some values of $s_{nl}$ and $l$ are forbidden and are not predicted by the method of Section \ref{sec:Summation};  $l=0$, $s=2$ for fermions  and, $l=1$, $s=3$ and $l=0$, $s=4$ for bosons are forbidden because the wavefunctions are 0 \cite{werner2008trapped}.
 
The interactions are enforced by the Bethe-Peierls boundary condition, Eq. (\ref{eq:BethePeierls}), and this allows for the values of $s_{nl}$ to be calculated, but only in the non-interacting and unitary regimes. In the intermediate regime this formalism breaks down. Note in Eq. (\ref{eq:HyperradialCondition}) $s_{nl}^2$ appears not $s_{nl}$. Hence there is a degree of arbitrarity in $s_{nl}$, convention is to choose $s_{nl}\in[0,\infty)$ for $s_{nl}^2\geq0$ and $s_{nl}\in i\cdot[0,\infty)$ for $s_{nl}^2<0$. The $s_{nl}^2<0$ case will be considered in more detail in Section \ref{sec:Efimov}.

In the non-interacting case Eq. (\ref{eq:BethePeierls}) implies $\varphi_{ls_{nl}}(0)=0$, and so the values of $s_{nl}$ are given \cite{liu2010three}
\begin{eqnarray}
{\rm NI\;fermions \;}s_{nl}=
\begin{cases}
2n+4 & l=0\\
2n+l+2 & l>0
\end{cases},
\\
{\rm NI\;bosons \;}s_{nl}=
\begin{cases}
2 & l=0\\
2n+6 & l=0\\
2n+l+4 & l=1\\
2n+l+2 & l>1
\end{cases},
\end{eqnarray}
where $n\in\mathbb{Z}_{\geq0}$.

In the unitary limit $s_{nl}$ will solve the following transcendental equation
\begin{eqnarray}
&&0=\nonumber\\
&&\frac{d\varphi_{ls_{nl}}}{d\alpha}\Big |_{\alpha=0}+\eta(-1)^l\frac{(1+\kappa)^2}{\kappa\sqrt{1+2\kappa}}\varphi_{ls_{nl}}\left(\arctan(\frac{\sqrt{1+2\kappa}}{\kappa})\right).\nonumber\\\label{eq:Transcendental}
\end{eqnarray}
See Table \ref{tab:sEigenvalues} for solutions to Eq. (\ref{eq:Transcendental}).
\\
\begin{table}
\begin{tabular}{|c|c|c|c|c|}
\hline
\multicolumn{2}{|c|}{} & 3 bosons & 2+1 fermions $\kappa=1$ & $\kappa=13.75$\\
\hline
$l$ & $n$ & \multicolumn{3}{c|}{$s_{nl}$} \\
\hline
\multirow{4}{4em}{0} & 0 & $i\cdot 1.006$\dots & 2.166\dots & 3.538\dots \\\cline{2-5}
& 1 & 4.465\dots & 5.127\dots & 4.802\dots \\\cline{2-5}
& 2 & 6.818\dots & 7.114\dots & 6.715\dots \\\cline{2-5}
& 3 & 9.324\dots & 8.832\dots & 10.912\dots \\
\hline
\multirow{4}{4em}{1} & 0 & 2.863\dots & 1.77\dots & $i\cdot 0.165$\dots  \\\cline{2-5}
& 1 & 6.462\dots & 4.358\dots & 3.940\dots \\\cline{2-5}
& 2 & 7.852\dots & 5.716\dots & 6.132\dots \\\cline{2-5}
& 3 & 9.822\dots & 8.053\dots & 8.211\dots \\
\hline
\multirow{4}{4em}{2} & 0 & 2.823\dots & 3.104\dots & 3.853\dots \\\cline{2-5}
& 1 & 5.508\dots & 4.795\dots & 4.965\dots \\\cline{2-5}
& 2 & 6.449\dots & 7.238\dots & 6.707\dots \\\cline{2-5}
& 3 & 9.272\dots & 8.837\dots & 8.782\dots \\
\hline
\multirow{4}{4em}{3} & 0 & 4.090\dots & 3.959\dots & 3.383\dots \\\cline{2-5}
& 1 & 5.771\dots & 6.127\dots & 6.062\dots \\\cline{2-5}
& 2 & 8.406\dots & 7.816\dots & 8.196\dots \\\cline{2-5}
& 3 & 9.607\dots & 10.172\dots & 10.200\dots \\
\hline
\end{tabular}
\caption{Three-body wavefunction eigenvalues, $s_{nl}$, at unitarity for the bosonic case and the fermionic case with $\kappa=1$ and $\kappa=13.75$, to three decimal places. }
\label{tab:sEigenvalues}
\end{table}
\\
In the $\kappa\rightarrow0$ limit, with $\eta=-1$ (fermions), Eq. (\ref{eq:Transcendental}) reduces to \begin{eqnarray}
0=\frac{d\varphi_{ls_{nl}}}{d\alpha}\Big |_{\alpha=0}-\delta_{l,0},
\end{eqnarray}
in turn this implies, for $\kappa \rightarrow0$,
\begin{eqnarray}
s_{nl}=
\begin{cases}
4n+2 & l=0\\
2n+l+1 & l>0
\end{cases},
\quad
\label{eq:kappa0energies}
\end{eqnarray}
where $n\in\mathbb{Z}_{\geq0}$. Note $l=0$, $s=2$ is normally forbidden but because $s\rightarrow2$ as $\kappa\rightarrow0$ this is never violated for finite mass imbalance. In the $\kappa\rightarrow\infty$ limit the right-hand term in Eq. (\ref{eq:Transcendental}) diverges unless $s_{nl}$ takes specific values that cause the $\varphi_{ls_{nl}}$ part of the term to be 0. This implies, for $\kappa\rightarrow\infty$
\begin{eqnarray}
s_{nl}=
\begin{cases}
2n+4 & l=0\\
2n+l+2 & l>0
\end{cases},
\quad
\end{eqnarray}
where $n\in\mathbb{Z}_{\geq0}$. 

The energies at unitarity are plotted in Figs. \ref{fig:FermionESpec}, \ref{fig:BosonESpec} and \ref{fig:ESpeck>1}. In Fig. \ref{fig:FermionESpec} the purple pluses indicate unitary energies for the $\kappa=1$ case and the cyan crosses for the $\kappa\rightarrow0$ case. In Fig. \ref{fig:BosonESpec} the black pluses indicate the unitary energies for $l=0$ and the orange crosses for $l=1$. In Fig. \ref{fig:ESpeck>1} the purple pluses indicate unitary energies for the $\kappa=1$ case and the cyan crosses for $\kappa=13.75$.

The energies calculated using the general method of Section \ref{sec:Summation} are in excellent agreement with the hyperspherical approach of this section. However, as noted in the captions of Figs. \ref{fig:BosonESpec} and \ref{fig:ESpeck>1}, some states are not marked at unitarity with an energy from a hyperspherical calculation. These states are Efimov states and they correspond to the case where $s_{nl}$ is imaginary. In this case Eq. (\ref{eq:HyperradialWavefunc}) is not the correct wavefunction and so $E_{\rm rel}=(2q+s_{nl}+1)\hbar\omega$ is not applicable.

\subsection{Efimov states}
\label{sec:Efimov}
The previous two sections have investigated the agreement between two different methods of calculating the unitary energy spectrum of trapped three-body systems including for bosons and the effects of mass imbalance. We find that the two methods are in excellent agreement, however, as noted in Figs. \ref{fig:BosonESpec} and \ref{fig:ESpeck>1}, there are some unitary energies predicted by the method of Section \ref{sec:Summation} that are not predicted by Eqs. (\ref{eq:HyperEnergy}) and (\ref{eq:Transcendental}). These unmarked states, associated with imaginary values of $s_{nl}$, are Efimov states, e.g. $s_{00}$ for bosons and $s_{01}$ for $\kappa=13.75$ fermions. 

Eq. (\ref{eq:HyperradialCondition}) is the Schr\"odinger equation for a particle moving in a two dimensions with two potential terms, a harmonic potential and a term proportional to $s_{nl}^2/R^2$. If $s_{nl}$ is purely imaginary then this potential is \textit{attractive} at short distances rather than repulsive. This necessitates a different class of solution compared to the case where $s_{nl}$ is purely real and these solutions are Efimov states. Physically speaking the Efimov effect is short-range interparticle interactions giving rise to an effective attractive long-range interaction due to a third particle mediating the effective interaction between the other two.

These states were first studied in Ref. \cite{efimov1971bound} in the free context and Ref. \cite{jonsell2002universal} in a trapped context. Efimov states were first observed in an ultracold gas in Ref. \cite{kraemer2006evidence}.

\begin{figure}[H]
\includegraphics[height=5.5cm,width=8.5cm]{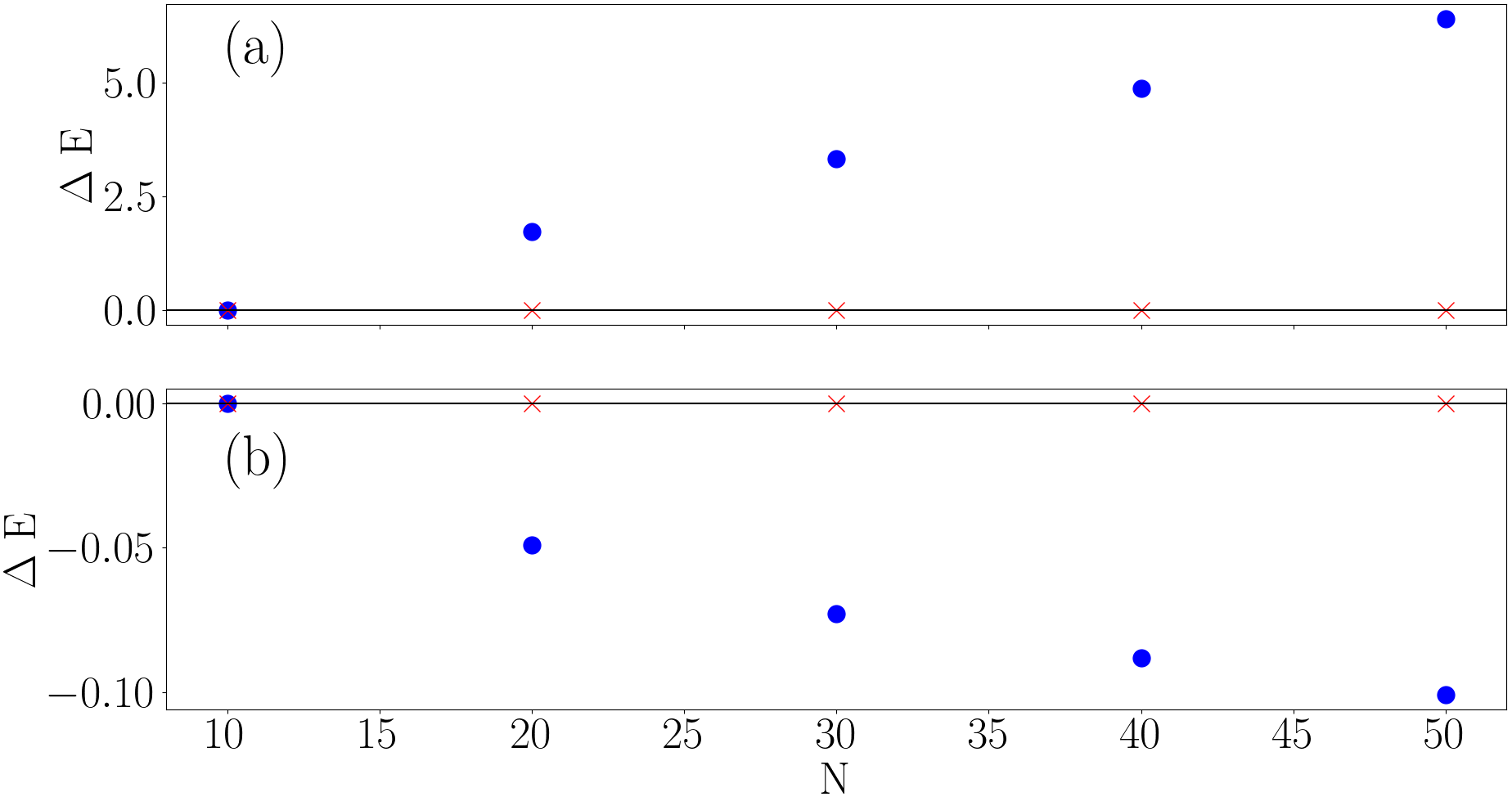}
\caption{The relative change in the energies of the lowest energy Efimov (blue circles) and lowest energy universal (red crosses) states calculated as per Section \ref{sec:Summation} with increasing matrix size in Eq. (\ref{eq:Matrix}). $N$ is the dimension of the matrix, i.e. the matrix is $N\times N$, and $\Delta E=(E-E_{N=10})/E_{N=10}$ is the proportional deviation away from the $N=10$ energy (universal or Efimov as appropriate). Panel (a) is the bosonic $l=0$ case ($E_{N=10}=-0.563\hbar\omega$) and the panel (b) is the fermionic $l=1$, $\kappa=13.75$ case ($E_{N=10}=1.507\hbar\omega$). The horizontal black line defines $\Delta E$=0.
}
\label{fig:EfimovDivergence}
\end{figure}

For imaginary $s_{nl}$ the hyperradial solution is given by \cite{werner2006unitary}
\begin{eqnarray}
F(R)=\frac{a_{\mu}}{R}W_{\dfrac{E_{\rm rel}}{2\hbar\omega},\dfrac{s_{nl}}{2}}\left(\frac{R^2}{a_{\mu}^2}\right),
\label{eq:HyperradialWhittaker}
\end{eqnarray}
where $W_{a,b}(x)$ is the Whittaker function. In the universal ($s_{nl}^2 \geq 0$) case Eqs. (\ref{eq:HyperradialCondition}) and (\ref{eq:HyperradialWavefunc}) imply $E_{\rm rel}=(2q+s_{nl}+1)\hbar\omega$. However, in the Efimov case the relative energy is left a free parameter. Using the method of Section \ref{sec:Summation} the energies of the Efimov states do not converge to a constant value as the matrix size is increased, whereas the universal states do converge. This non-convergence of the Efimov states is shown in Fig. \ref{fig:EfimovDivergence}, where the change in energy, $\Delta E=(E-E_{N=10})/E_{N=10}$ is plotted as a function of the matrix size $(N)$ in Eq. (\ref{eq:Matrix}). Figs. \ref{fig:EfimovDivergence}$(\rm a)$ and \ref{fig:EfimovDivergence}$(\rm b)$ plot $\Delta E$ for bosons and fermions respectively. From these plots it is clear that as $N$ is increased the energy for the universal states ($s_{nl}^2>0$) states remains constant, whereas the Efimov state energies diverge. This divergence arises from the fact that the Hamiltonian and Bethe-Peierls boundary condition do not contain enough information to uniquely specify the energies of the Efimov states. We find that the Efimov energies diverge logarithmically with $N$. As such an additional boundary condition is needed to fix the Efimov energies. 

The Efimov hyperradial function, Eq. (\ref{eq:HyperradialWhittaker}), oscillates increasingly rapidly as $R/a_{\mu}\rightarrow0$ (but remains bounded). To find the Efimov states we require the phase of the oscillation to be fixed, and impose the boundary condition 
\begin{eqnarray}
F(R) \underset{R\rightarrow0}{=} A\sin\Big[|s_{0l}|\ln(\frac{R}{a_{\mu}})-|s_{0l}|\ln(\frac{R_{t}}{a_{\mu}}) \Big],
\label{eq:EfimovBC}
\end{eqnarray}
where $A$ is a constant and $R_{t}>0$ is the three-body parameter. The short range behaviour of the Efimov hyperradial wavefunction is given by 
\begin{eqnarray}
\dfrac{a_{\mu}}{R}W_{\dfrac{E}{2\hbar\omega},\dfrac{s_{0l}}{2}}&&\left(\dfrac{R^2}{a_{\mu}^2}\right)\underset{R\rightarrow0}{=}\nonumber\\
&& \dfrac{\Gamma[-s_{0l}]}{\Gamma\left[\dfrac{1-s_{0l}-E/\hbar\omega}{2}\right]}\left(\dfrac{R}{a_{\mu}}\right)^{s_{0l}}\nonumber\\
&&+\dfrac{\Gamma[s_{0l}]}{\Gamma\left[\dfrac{1+s_{0l}-E/\hbar\omega}{2}\right]}\left(\dfrac{R}{a_{\mu}}\right)^{-s_{0l}}.\nonumber\\\label{eq:EfimovShortRange}
\end{eqnarray}
From this we can derive a transcendental equation that determines the energy spectrum as a function of $R_{t}$ \cite{jonsell2002universal}
\begin{eqnarray}
-|s_{0l}|\ln\left(\frac{R_{t}}{a_{\mu}}\right)&=&\arg\left\{ \frac{\Gamma\left[\dfrac{1+s_{0l}-E}{2}\right]}{\Gamma[1+s_{0l}]} \right\} \quad {\rm mod}\;\pi.\qquad
\label{eq:EfimovEnergy}
\end{eqnarray}
The energy spectrum is unbounded from above and below. We index the states with $q\in\mathbb{Z}$, and define the $q=0$ state to be the first positive energy state at $R_{t}/a_{\mu}=\exp(\pi/|s_{0l}|)$ with $q>0$ referring to higher energy states and $q<0$ to lower energy states. For example for $R_{t}/a_{\mu}=1$ and $s_{0l}=i\cdot 1.006$, the energies corresponding to $q=3,2,1,0,-1,-2$ are given $E_{\rm rel}/\hbar\omega\approx 6.60, 4.48, 2.27,-0.85, -566, -291 649$ respectively and at $R_{t}/a_{\mu}=\exp(\pi/|s_{0l}|)$ the energies corresponding to $q=3,2,1,0,-1,-2$ are given $E_{\rm rel}/\hbar\omega\approx 8.686, 6.60, 4.48, 2.27, -0.85,-566 $ respectively. In general the energy of the $q=N$ state at $R_{t}/a_{\mu}=\exp(\pi/|s_{0l}|)$ is equal to the energy at $R_{t}/a_{\mu}=1$ of the $q=N+1$ state. For fixed $R_{t}$ the states with $q\geq0$ are have a regular spacing of $\approx 2\hbar\omega$ whereas the spacing of the $q<0$ states grows larger for more negative energies.

\begin{figure}[H]
\includegraphics[height=5.5cm,width=8.5cm]{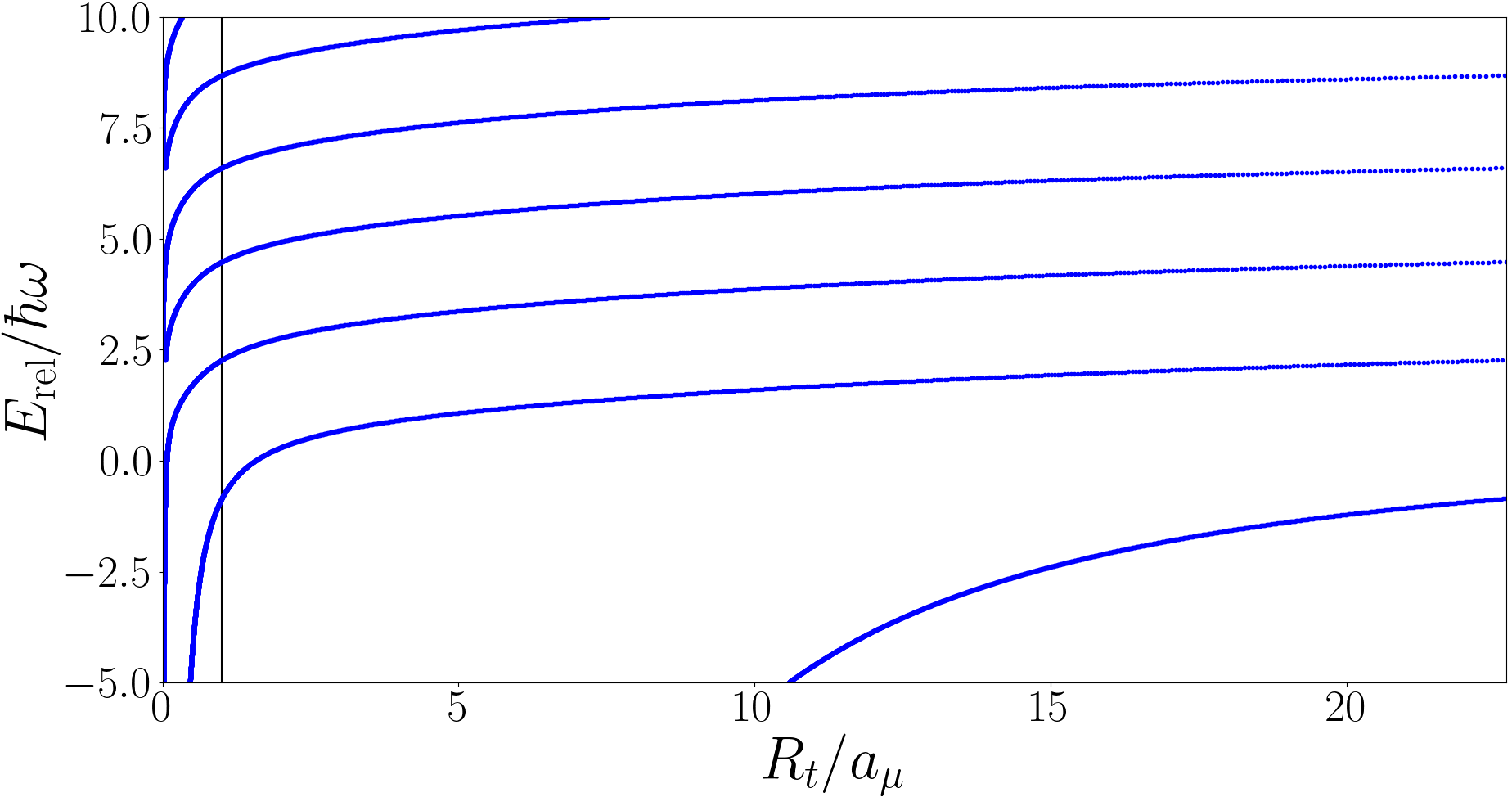}
\caption{The energy spectrum for Efimov states as defined by Eq. (\ref{eq:EfimovEnergy}). Calculated using $s=i\cdot 1.006$. The upper limit on the horizontal axis is $e^{\pi/|s|}\approx22.7$, and the vertical black line is $R_{t}=a_{\mu}$.}
\label{fig:EfimovEspec}
\end{figure}

Physically speaking the additional parameter $R_{t}$ is needed because in the Efimov states the specifics of the interparticle interaction become significant, due to the attractive $s_{0l}^2/R^2$ potential term, and can no longer accurately be considered a contact interaction. $R_{t}$ describes the short-range (but not zero-range) properties of the interparticle interaction and will, in general, differ for different species of atoms although some work suggests that it will vary little in units of the van der Waals length in practice \cite{wang2012origin}.

Efimov states also exist in the fermionic case for sufficient mass imbalance \cite{efimov1973energy,kartavtsev2007low}. By solving Eq. (\ref{eq:Transcendental}) for $\kappa$, with $s_{nl}=0$ and $\eta=-1$, we find the mass imbalance for which $s_{nl}$ becomes imaginary and Efimov states are then allowed. There are no Efimov states for $l$ is even. They appear for $l=1,3,5\dots$ for $\kappa\gtrsim13.606\dots,75.994\dots,187.958\dots$ respectively. This is in good agreement with previous work \citep{kartavtsev2007low, kartavtsev2008universal, endo2011universal, petrov2012few}. Using the method of Section \ref{sec:Summation}, the Efimov states appear (or do not appear as appropriate) at these predicted values of $l$ and $\kappa$. Additionally while the real values of $s_{nl}$ converge in the $\kappa\rightarrow\infty$ limit the imaginary solutions to Eq. (\ref{eq:Transcendental}) do not converge.

As described above; in the method of Section \ref{sec:Summation} the Efimov energies depend on matrix size and are divergent with increasing matrix size, and in the approach of Section \ref{sec:Hyperspherical} the Efimov energies are not fixed so an additional parameter is required. In fact these two methods, the matrix size and the three-body parameter, are closely linked; for every finite matrix size there is a value of $R_{t}$ that produces the same unitary Efimov energy spectrum to a high degree of accuracy. This result implies that the boundary condition Eq. (\ref{eq:EfimovBC}) is enforced by the finite matrix size of Eq. (\ref{eq:Matrix}). The errors between the Efimov energies calculated with the two methods are given in Tables \ref{tab:Bosons} and \ref{tab:Fermions}. In Tables \ref{tab:Bosons} and \ref{tab:Fermions} the value of $R_{t}$ is calculated by substituting the lowest Efimov energy calculated with the matrix method for a given matrix size	 $N$ into Eq. (\ref{eq:EfimovEnergy}), which corresponds to either $q=0$ or $q=-1$ depending on $N$ and the particle symmetry. The error between the two methods of calculating the Efimov energies is $\lesssim 1\%$ for all calculated values and is most likely due to the linear interpolation required to determine the unitary energies using the method of Section \ref{sec:Summation}. 

However away from unitarity this equivalence no longer holds. For a general value of $a_{\rm s}$, unlike at unitarity, there is no value of $s_{nl}$ such that there exists a value of $R_t$ that matches the non-unitary Efimov energy spectrum of Section \ref{sec:Summation}. Repeating the above comparison between the Efimov energy spectra (for the same values of $s_{nl}$) for bosons at $a_{\rm m}/a_{\rm s}=-0.25$ instead of at unitarity lead to errors of $\approx 2\%$ compared to $\lesssim1\%$ at unitarity. The errors grow as distance from unitarity grows, with an error of $\approx 13\%$ at $a_{\rm m}/a_{\rm s}=-5$. In the fermionic case we have similar results with errors increasing as one moves further from unitarity. The equivalence breaks down in the intermediate $a_{\rm s}$ regime because $s_{nl}$ is no longer well defined. Away from the unitary or non-interacting regimes the left-hand side of Eq. (\ref{eq:Transcendental}) depends on $\rho$ and Eq. (\ref{eq:HyperForm}) cannot be used to describe the three-body relative wavefunction. As Eq. (\ref{eq:EfimovEnergy}) depends on $s_{nl}$ it cannot be used to create an energy spectrum in the intermediate $a_{\rm s}$ regime.

In drawing an equivalence between these two methods of calculating the unitary Efimov energy spectrum we must note that the spectrum determined by Eq. (\ref{eq:EfimovEnergy}) is unbounded above and below whereas the Efimov energy spectrum defined by Eq. (\ref{eq:Matrix}) is unbounded above but not below. In Figs. \ref{fig:BosonESpec} and \ref{fig:ESpeck>1} one can see that the Efimov energies approach non-interacting energies in the $a_{\rm s}^{-1}\rightarrow -\infty$ limit. In order to preserve the appropriate multiplicity of the non-interacting eigenstates, the lowest observed energy states must in fact be the lowest energy states. That being said, the lower energies predicted by Eq. (\ref{eq:EfimovEnergy}) are significantly lower, e.g. for $R_{t}=11.555$ the $q=-1$ state has $E\approx-4.164\hbar\omega$ as per Table \ref{tab:Bosons} but for $q=-2$ we have $E\approx-566\hbar\omega$, and this makes direct confirmation of the absence of these states in the formulation of Eq. (\ref{eq:Matrix}) numerically challenging.

In the hyperspherical formulation Eqs. (\ref{eq:HyperangularWavefunc}) and (\ref{eq:HyperradialWavefunc}) define an orthogonal set of states in both the unitary and non-interacting regimes. Integrating over the universal hyperradial wavefunction gives an orthogonality in $q$ and integrating over the hyperangular part gives an orthogonality in $s_{nl}$ \cite{fedorov1993efimov}. The Efimov hyperradial wavefunctions are orthogonal with the energy spectrum defined by Eq. (\ref{eq:EfimovEnergy}).

The orthogonality of the Efimov wavefunctions is not obvious and here we provide a short proof. Consider some energy $E$ and some other energy $E'\neq E$ both of which satisfy Eq. (\ref{eq:EfimovEnergy}) for the same values of $R_{t}$ and $s_{nl}=s$. The overlap of two Efimov hyperradial wavefunctions is given by \cite{gradshteyn2014table, kerin2022effects}
\begin{widetext}
\begin{eqnarray}
\bra{F_{q's}}\ket{F_{qs}}=&&\frac{a_{\mu}^2}{2}\frac{\Gamma\left(s+1\right)\Gamma(-s)}{\Gamma\left(\dfrac{1-E/\hbar\omega-s}{2}\right)\Gamma\left(\dfrac{3-E'/\hbar\omega+s}{2}\right)}{}_{2}F_{1}\bigg(1,\frac{1-E/\hbar\omega+s}{2}; \frac{3-E'/\hbar\omega+s}{2};1\bigg)\nonumber\\
+&&\frac{a_{\mu}^2}{2}\frac{\Gamma\left(1-s\right)\Gamma(s)}{\Gamma\left(\dfrac{1-E/\hbar\omega+s}{2}\right)\Gamma\left(\dfrac{3-E'/\hbar\omega-s}{2}\right)}
{}_{2}F_{1}\bigg(1,\frac{1-E/\hbar\omega-s}{2};\frac{3-E'/\hbar\omega-s}{2};1\bigg),
\label{eq:EfimovOverlap}
\end{eqnarray}
\end{widetext}
where $F_{q's}$ has energy $E'$ and $F_{qs}$ has energy $E$. Note the identity
\begin{eqnarray}
{}_{2}F_{1}(a,b;c;1)=\frac{\Gamma\left( c \right) \Gamma\left( c-a-b \right)}{\Gamma\left( c-a \right) \Gamma\left( c-b \right)}, \; {\rm if} \; \mathbb{R}(c) > \mathbb{R}(a+b),\nonumber\\
\end{eqnarray}
which applies if $E'<E$ which we can always set to be true given that $\bra{F_{q's}}\ket{F_{qs}}=\bra{F_{qs}}\ket{F_{q's}}$. Additionally note $\Gamma(1+z)=z\Gamma(z)$ and that the two terms in Eq. (\ref{eq:EfimovOverlap}) are complex conjugates. We then have
\begin{eqnarray}
&&\bra{F_{q's}(R)}\ket{F_{qs}(R)}=a_{\mu}^2 \times \nonumber\\
&& \Re \left[ \frac{2\hbar\omega }{s(E'-E)} \frac{\Gamma(1-s)}{\Gamma\left(\dfrac{1-E/\hbar\omega-s}{2}\right)}
\frac{\Gamma(1+s)}{\Gamma\left(\dfrac{1-E'/\hbar\omega+s}{2}\right)} \right].\nonumber\\
\label{eq:EfimovOverlapReduced}
\end{eqnarray}
Note that the second and third terms in Eq. (\ref{eq:EfimovOverlapReduced}) (or their complex conjugates) appear in the right-hand side of Eq. (\ref{eq:EfimovEnergy}) and therefore have opposite phase (up to modulo $\pi$). We can then write
\begin{eqnarray}
Be^{-iA}&=\dfrac{\Gamma(1-s)}{\Gamma\left(\dfrac{1-E/\hbar\omega-s}{2}\right)},\\
Ce^{i(A+n\pi)}&=\dfrac{\Gamma(1+s)}{\Gamma\left(\dfrac{1-E'/\hbar\omega+s}{2}\right)},
\end{eqnarray}
where $A,B,C\in\mathbb{R}$ and $n\in\mathbb{Z}$. As such in Eq. (\ref{eq:EfimovOverlapReduced}) the first term in the square brackets is purely imaginary and the product of the second and third terms is purely real, hence $\bra{F_{q's}(R)}\ket{F_{qs}(R)}=0$. Since $E$ and $E'$ are only required to be not equal and satisfy Eq. (\ref{eq:EfimovEnergy}) for the same $R_t$ and $s_{nl}$ we have that Eq. (\ref{eq:EfimovEnergy}) produces an orthogonal spectrum of energies.

The wavefunctions given by Eq. (\ref{eq:psi1}) and (\ref{eq:HyperForm}) both solve the three-body Hamiltonian and satisfy the Bethe-Peierls condition and reproduce the same energy spectrum in the non-interacting and unitary regimes. These are different representations of the same wavefunction and as selecting a specific $R_t$ defines an orthogonal subset of Efimov energies a specific matrix size also selects an orthogonal subset.

\begin{table}

\begin{tabular}{|c|c|c|c|c|c|}
\hline
Bosons & $q=0$ & $q=1$ & $q=2$ & $q=3$ & $q=4$ \\
\hline
$N=10$ & -0.563 & 2.393 & 4.612 & 6.747 & 8.849\\
\hline
$R_{t}/a_{\mu}=1.131$ & -0.563 & 2.365 & 4.566 & 6.682 & 8.763\\
\hline
Error \% & 0 & 1.191 & 1.000 & 0.963 & 0.967\\
\hline
 & $q=-1$ & $q=0$ & $q=1$ & $q=2$ & $q=3$ \\
\hline
$N=20$ & -1.531 & 2.112 & 4.353 & 6.490 & 8.588\\
\hline
$R_{t}/a_{\mu}=18.216$ & -1.531 & 2.094 & 4.326 & 6.455 & 8.544\\
\hline
Error \% & 0 & 0.870 & 0.621 & 0.543 & 0.511\\
\hline
$N=30$ & -2.428 & 1.943 & 4.198 & 6.341 & 8.441\\
\hline
$R_{t}/a_{\mu}=14.897$ & -2.428 & 1.928 & 4.178 & 6.315 & 8.409\\
\hline
Error \% & 0 & 0.767 & 0.494 & 0.410 & 0.375 \\
\hline
$N=40$ & -3.301 & 1.823 & 4.087 & 6.234 & 8.336\\
\hline
$R_{t}/a_{\mu}=12.912$ & -3.301 & 1.810 & 4.070 & 6.212 & 8.310\\
\hline
Error \% & 0 & 0.709 & 0.423 & 0.344 & 0.310\\
\hline
$N=50$ & -4.164 & 1.731 & 4.001 & 6.150 & 8.253\\
\hline
$R_{t}/a_{\mu}=11.555$ & -4.164 & 1.719 & 3.990 & 6.131 & 8.231\\
\hline
Error \% & 0 & 0.681 & 0.386 & 0.303 & 0.266\\
\hline
\end{tabular}
\caption{Comparison of the matrix and three-body parameter Efimov energies (in units of $\hbar\omega$) for bosons. The three-body parameter is found by substituting the lowest unitary Efimov energy predicted by the matrix method into Eq. (\ref{eq:EfimovEnergy}). For matrix size $N=10$ this means that the $q=0$ state is the corresponding lowest energy state, and for $N\geq20$ $q=-1$ is the corresponding state. All values are stated to 3 decimal places. Errors are $\lesssim 1\%$, with the primary source of error being the accurate calculation of the matrix energies, this also affects the accuracy of $R_{t}$ as $R_{t}$ is chosen depending on the lowest Efimov state energy.}
\label{tab:Bosons}
\end{table}

\begin{table}

\begin{tabular}{|c|c|c|c|c|c|}
\hline
Fermions & $q=0$ & $q=1$ & $q=2$ & $q=3$ & $q=4$ \\
\hline
$N=10$ & 1.507 & 3.626 & 5.713 & 7.787 & 9.857\\
\hline
$R_{t}/a_{\mu}=2.59\times10^{7}$ & 1.507 & 3.611 & 5.680 & 7.731 & 9.773\\
\hline
Error \% & 0 & 0.380 & 0.573 & 0.721 & 0.8544\\
\hline
$N=20$ & 1.433 & 3.525 & 5.588 & 7.640 & 9.685\\
\hline
$R_{t}/a_{\mu}=1.84\times10^{7}$ & 1.43 & 3.51 & 5.57 & 7.61 & 9.65\\
\hline
Error \% & 0 & 0.150 & 0.221 & 0.272 & 0.312\\
\hline
$N=30$ & 1.397 & 3.477 & 5.531 & 7.574 & 9.611\\
\hline
$R_{t}/a_{\mu}=1.50\times10^{7}$ & 1.397 & 3.474 & 5.524 & 7.563 & 9.594\\
\hline
Error \% & 0 & 0.085 & 0.127 & 0.153 & 0.178\\
\hline
$N=40$ & 1.374 & 3.447 & 5.496 & 7.535 & 9.567\\
\hline
$R_{t}/a_{\mu}=1.30\times10^{7}$ & 1.374 & 3.445 & 5.491 & 7.526 & 9.556\\
\hline
Error \% & 0 & 0.061 & 0.087 & 0.106 & 0.121\\
\hline
$N=50$ & 1.355 & 3.425 & 5.463 & 7.491 & 9.515\\
\hline
$R_{t}/a_{\mu}=1.14\times10^{7}$  & 1.355 & 3.421 & 5.463 & 7.496 & 9.523\\
\hline
Error \% & 0  & 0.113 & 0.005 & 0.063 & 0.082\\
\hline
\end{tabular}
\caption{Comparison of the matrix and three-body parameter Efimov energies (in units of $\hbar\omega$) for fermions with $\kappa=13.75$ and $l=1$. The three-body parameter is found by substituting the lowest unitary Efimov energy predicted by the matrix method into Eq. (\ref{eq:EfimovEnergy}). All values are stated to 3 decimal places. Errors are $\lesssim 1\%$, with the primary source of error being the accurate calculation of the matrix energies, this also affects the accuracy of $R_{t}$ as $R_{t}$ is chosen depending on the lowest Efimov state energy.}
\label{tab:Fermions}
\end{table}

\section{Conclusion}
\label{sec:Conc}
In this work we have considered exact solutions for three particles in a spherically symmetric trap. We have provided results for both fermionic systems with mass imbalance and bosonic systems where by definition there is no mass imbalance. In each case these results can be used to investigate physical phenomena such as quench dynamics \cite{PhysRevA.86.053631,PhysRevLett.102.160401,liu2010three,Cui2012,PhysRevLett.96.030401,PhysRevA.85.033634,PhysRevLett.107.030601,mulkerin2012universality,nascimbene2010exploring,Science335_2010,
levinsen2017universality} and the derivation of the equation of state of such a gas \cite{bougas2020stationary,bougas2019analytical,budewig2019quench, kehrberger2018quantum}. 

We have also revisited, for fermions with mass imbalance, and bosons, the Efimov states in such systems. Specifically we have compared solutions with the matrix approach, Section \ref{sec:Summation}, and the hyperspherical approach, Section \ref{sec:Hyperspherical}. We find in each case very good agreement between the two approaches, at unitarity, for fermions with and without mass imbalance and for mass-balanced bosons. Finally, we have investigated, in Section \ref{sec:Efimov} the emergence of Efimov states for the mass-imbalanced fermion case and the mass-balanced boson case. Remarkably we find a connection between the matrix and three-body parameter approaches in determining the energy of Efimov states at unitarity. Specifically, for the matrix approach we find the energy of an Efimov state diverges with increasing matrix size, see Fig. \ref{fig:EfimovDivergence}. However for every matrix size there is a value of $R_t$ that produces the same energy spectrum, see Tables \ref{tab:Bosons} and \ref{tab:Fermions}.

This parameterisation of the energy spectrum in the intermediate regime lays the groundwork for future calculations regarding three-body quench dynamics, virial coefficients and Tan contacts. Three-body quench dynamics calculations can currently be performed for the non-interacting to unitary (and vice-versa) quenches using the hyperspherical formalism \cite{kerin2022quench, kerin2022effects}, but calculations for intermediate quenches are lacking. In this work we have uniquely specified the energies and wavefunctions in the intermediate regime, the remaining barrier to calculating for an arbitrary quench is the difficulty of calculating the wavefunction overlaps using Eq. (\ref{eq:psi1}), correctly accounting for the permutation operator is quite difficult. However in the hyperspherical formalism, Eq. (\ref{eq:HyperForm}), integrating over the terms acted on by the permutation operator can be done by utilising a ``kinematic rotation'', a coordinate transform on the hyperangle $\alpha$ which is a function of $r$ and $\rho$ \cite{nielsen2001three, braaten2006universality,thogersen2009universality, fedorov2001regularization, fedorov1993efimov}. It should be possible to modify the ``kinematic rotation'' technique and, with the wavefunctions and energies now specified for general $a_{\rm s}$, calculate quench observables for a general quench between intermediate scattering lengths.

\section{Data Availability Statement}

All data will be provided upon request to A. D. Kerin.

\section{Acknowledgements}
 A.D.K. is supported by an Australian Government Research Training Program Scholarship and by the University of Melbourne.
 
 With thanks to Victor Colussi for enlightening conversations.

\bibliographystyle{apsrev4-1}{}
\bibliography{Few-Body-Refs}


\end{document}